\documentclass[twoside,11pt]{article}

\usepackage[final]{melba}

\usepackage{amsmath,amsfonts}

\melbaheading{1}{https://www.melba-journal.org/article/18135-an-uncertainty-driven-gcn-refinement-strategy-for-organ-segmentation}{2020}{1-27}{09/2020}{11/2020}{Soberanis-Mukul and Navab and Albarqouni}{Medical Imaging with Deep Learning (MIDL) 2020}{Marleen de Bruijne, Tal Arbel, Ismail Ben Ayed, Hervé Lombaert}

\ShortHeadings{Uncertiny-driven GCN refinement for organ segmentation}{Soberanis-Mukul, Navab, Albarqouni}
\firstpageno{1}

\title{An Uncertainty-Driven GCN Refinement Strategy \\for Organ Segmentation}

\author{\name Roger D. Soberanis-Mukul \email roger.soberanis@tum.de \\  
	\addr Computer Aided Medical Procedures, Technical University of Munich, Germany
	\AND
	\name Nassir Navab \email nassir.navab@tum.de \\
	\addr Computer Aided Medical Procedures, Technical University of Munich, Germany \\
	Computer Aided Medical Procedures, Johns Hopkins University, Baltimore, USA
	\AND
	\name Shadi Albarqouni \email shadi.albarqouni@tum.de \\
	\addr Computer Aided Medical Procedures, Technical University of Munich, Germany \\
    Helmholtz AI, Helmholtz Center Munich, Germany
}

\begin{document}

\maketitle

\begin{abstract}
Organ segmentation in CT volumes is an important pre-processing step in many computer assisted intervention and diagnosis methods. In recent years, convolutional neural networks have dominated the state of the art in this task. However, since this problem presents a challenging environment due to high variability in the organ's shape and similarity between tissues, the generation of false negative and false positive regions in the output segmentation is a common issue. Recent works have shown that the uncertainty analysis of the model can provide us with useful information about potential errors in the segmentation. In this context, we proposed a segmentation refinement method based on uncertainty analysis and graph convolutional networks. We employ the uncertainty levels of the convolutional network in a particular input volume to formulate a semi-supervised graph learning problem that is solved by training a graph convolutional network. To test our method we refine the initial output of a 2D U-Net. We validate our framework with the NIH pancreas dataset and the spleen dataset of the medical segmentation decathlon. We show that our method outperforms the state-of-the-art CRF refinement method by improving the dice score by 1\% for the pancreas and 2\% for spleen, with respect to the original U-Net's prediction. Finally, we perform a sensitivity analysis on the parameters of our proposal and discuss the applicability to other CNN architectures, the results, and current limitations of the model for future work in this research direction.
For reproducibility purposes, we make our code publicly available at~\url{https://github.com/rodsom22/gcn\_refinement}.
\end{abstract}

\begin{keywords}
  Organ segmentation refinement, Uncertainty Quantification, Graph Convolutional Networks, Semi-Supervised Learning
\end{keywords}

\section{Introduction}
    In recent years, deep convolutional neural networks (CNN)  have become the standard in different learning problems in computer vision, like classification, localization, and segmentation. Inspired by this development, researchers have proposed CNN architectures for the processing of medical images, in different modalities. Image segmentation is a common task addressed by both the computer vision and the medical image processing communities. Segmentation of medical structures is an initial step in many computer-aided procedures, like computer-assisted navigation and detection. However, the necessity of experts for training-example annotation, the similarity in tissues,  and the inter-patient variation of anatomical structures add additional challenges compared with real-world images, leading to potential errors in the CNN predictions. 

\begin{figure}[t]
\begin{center}
\includegraphics[width=0.69\textwidth]{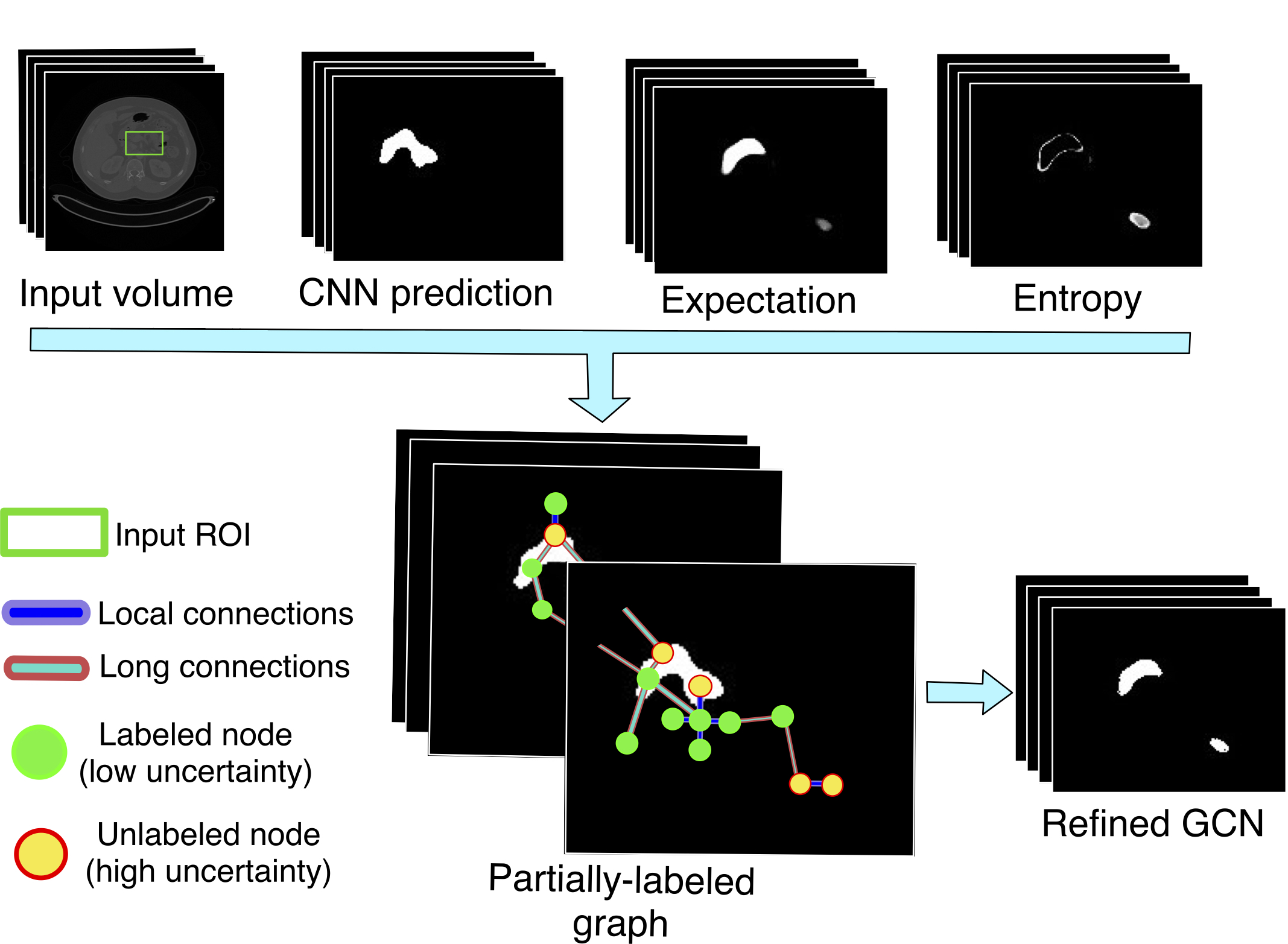}
\end{center}
\caption{A graphical overview of our GCN refinement strategy. Our proposal employs the input's intensities, and CNN's prediction, expectation, and entropy (uncertainty) to formulate the refinement problem as a semi-supervised graph learning problem.} \label{fig:graphic_overview}
\end{figure}

Organ segmentation in CT has been a topic of research. Recent CNN architectures employ single or multiple aggregations of CNN models, using two-dimensional \citep{bib:Zhou17} or three-dimensional networks \citep{bib:Zhu17,bib:Roth18}. Works that include shape and geometric priors have also been proposed \citep{bib:yuyin19,bib:yao19,degel2018domain}. A common practice to improve the segmentation of a model is the inclusion of a post-processing refinement step, included after the inference process of the CNN.  Methods based on conditional random fields (CRF) \citep{bib:Krahenbuhl11} are examples of a refinement strategy. Even though refinement methods can be the final step of the segmentation process, it can also serve as an intermediate step for improving model performance. For example, \cite{bib:Wang18} use a CRF-based method to generate a set of scribbles. In combination with user-defined scribbles, the results are used to perform an image specific fine-tune of a CNN. Similarly, semi-supervised learning methods can use refined predictions as pseudo-labels to allow including unlabeled data in CNN's training process \citep{bib:Bai17}. The work in \citep{bib:li} addresses the problem of loss in spatial correlation in the task of metastasis detection in Whole-slide images, due to the subdivision of the image in independent patches. The authors propose an architecture composed of a CNN that processes a group of input patches with a CRF on top. This CRF is employed to consider neighborhood information in the classification task, bringing consistency in the model's prediction.

CRF employs CNN prediction together with spatial and intensity similarity between the pixels in CT slices to refine the segmentation. In this sense, additional information regarding the correctness of the prediction could bring useful information to the process. Related to this idea, Gal \citep{bib:Kendall17} shows that a stochastic Gaussian process can be approximated through the dropout layers of regular CNNs, in a process known as Monte Carlo dropout (MCDO). This brings the possibility to estimate the uncertainty of recent CNN segmentation models. CNN uncertainty has proved to be useful as an attention mechanism in semi-supervised learning \citep{bib:Xia18}. Recent works in computer vision have started to explore its capabilities for finding potential misclassified regions for segmentation refinement purposes \citep{bib:Dias18}. In the medical context, the ability of uncertainty to reflect incorrect predictions has been recently studied \citep{bib:Nair18}. Similarly, a recent work presented by \cite{bib:Yu19}, uses the uncertainty of a teacher model to select the pseudo-labels to train a student model. 

Since uncertainty can bring insights regarding the potential errors in the segmentation, we still need a way to incorporate this knowledge into a refinement pipeline. In this work, we propose a method to formulate the segmentation refinement problem of CT data as a semi-supervised graph learning problem, that is solved using graph convolutional neural networks (GCN). Graph representations of three-dimensional data have been applied for refinement \citep{bib:Kamnitsas16}, similarly, recent works have started to explore the application of graph convolutional networks (GCN) for the segmentation of tubular structures, like airways \citep{bib:selvan,bib:juarez} and vessels \citep{bib:shin}. In this work, we explore the use of recent GCNs with sparse graphs-representations of 3-D data for the organ segmentation refinement task.   

For a given CNN, we first apply MCDO to obtain the model's expectation and uncertainty, this last expressed by the model's entropy. This is used to divide the CNN output into high confidence points (background or foreground) and low confidence points. With this information, we define a semi-labeled graph that is used to train a GCN in a semi-supervised way using the high confidence predictions.  The refined segmentation is obtained by re-evaluating the full graph in the trained GCN (see Fig. \ref{fig:graphic_overview}).  To our best knowledge, this is the first time a semi-supervised GCN learning strategy is employed in the medical image segmentation task, specifically, for single organ segmentation. Also, this work presents one of the first cases of using GCN and uncertainty analysis for organ segmentation refinement. We perform experiments for refining the segmentation of a U-Net on CT data for the pancreas and spleen segmentation problems. We compare our results with the popular CRF refinement method showing a better improvement over the initial CNN prediction and CRF refinement.

This work presents an extension of our initial results presented in \cite{bib:soberanis20}. We have extend our initial analysis into a sensitivity analysis presented in section \ref{sec:sen_analisys}. We compare with an additional connectivity scheme in section \ref{sec:node_conn}. We also perform a thorough study on the components of our edge weighting function in sections \ref{sec:edge_weighting} to \ref{sec:div_only}. Finally, we evaluate the performance of the refinement strategy on a second CNN architecture, namely QuickNat~\citep{bib:roy2019quicknat,bib:roy2019bayesian}, in section \ref{sec:other_models}, and discuss some insights on 2D vs. 3D CNN architectures in section \ref{sec:3d_architectures}.  
\section{Methods}
    \textit{Overview}: Consider an input CT volume $V$ with $V(x)$ the intensity value at the voxel position $x\in \mathbb{R}^3$; consider also, a trained CNN $g(V(x); \theta)$ with parameters $\theta$; and a segmented volume $Y(x) = g(V(x); \theta)$ with $Y(x) \in \{0, 1\}$. Our objective, is to refine the segmentation $Y$ using a GCN trained on a graph representation of the input data. Our framework operates as a post-processing step (one volume at a time) and assumes that no information about the real segmentation (ground truth) is available.

We first look for a binary volume $U_b$ used to highlight the potential false positives and false negatives elements of $Y$. The second step uses $U_b$, together with information coming from $Y$, $g$, and $V$, to refine the segmentation $Y$. We use uncertainty analysis to define $U_b$. For the second step, we solve the refinement problem using a semi-supervised GCN trained on a graph representation of our input volume.

\subsection{Uncertainty Analysis: Finding Incorrect Elements}\label{subsec:unc_analysis}

Incorrect elements are estimated considering the uncertainty of $g$.  We employ MCDO approximation \citep{bib:Kendall17,bib:gal} to evaluate the uncertainty of the CNN. This strategy can be applied to any model trained with dropout layers, without modifying or retraining the model. This attribute makes it ideal for a post-processing refinement algorithm. \cite{bib:gal} showed that a neural network trained with dropout layers before the convolutional layers is equivalent to approximate the probabilistic deep Gaussian Process. Following MCDO, we use the dropout layers of the network in inference time, and perform $T$ stochastic passes on the network to get the expectation of the model's prediction:
\begin{equation}
\mathbb{E}(x) \approx {\frac{1}{T}}\sum_{t=1}^{T}g({V(x)}, \theta _t),
\end{equation}
with $\theta_t$ the model parameters after applying dropout in the pass $t$. The model uncertainty $\mathbb{U}$ is given by the entropy, computed as 
\begin{equation}
\mathbb{U}(x) = H(x) = -\sum_{c=1}^M P(x)^c\log{P(x)^c},
\end{equation}
with $P(x)^c$ being the true probability of the voxel $x$ to belong to class $c$, and $M$ is the number of classes ($M=2$ in our binary segmentation scenario). To approximate this probability, we use the expectation of the model's prediction $\mathbb{E}$. Finally, we define the potential incorrect elements by applying a binary threshold on the entropy volume: 
\begin{equation}
U_b(x) = \mathbb{U}(x) > \tau,
\end{equation}
where the uncertainty threshold $\tau$ controls the entropy necessary to consider a voxel $x \in Y$ as uncertain.

\subsection{Graph Learning for Segmentation Refinement}\label{gcn_refinemet}

At this point, we have a binary mask $U_b$ indicating voxels with high uncertainty. The uncertainty analysis only tells us that the model is not confident about its predictions.  Some of the elements indicated by $U_b$ could be indeed correct and its value should not be changed. 
However, we can use a learning model that trains on high confidence voxels to reclassify (refine) the output of the CNN $g$.
Using the information from the uncertainty analysis, we can define a partially-labeled graph, where the voxels are mapped to nodes, and neighborhood relationship to edges.  In this way, we formulate the refinement problem as a semi-supervised graph learning problem. We address this mapped problem by training a GCN on the high confidence voxels using the methods presented in \cite{bib:Kipf17}. The rest of this section describes the formulation of our partially-labeled graph. 

\subsubsection{Partially-Labeled Nodes}
\begin{figure}
\begin{center}
\includegraphics[width=0.8\textwidth]{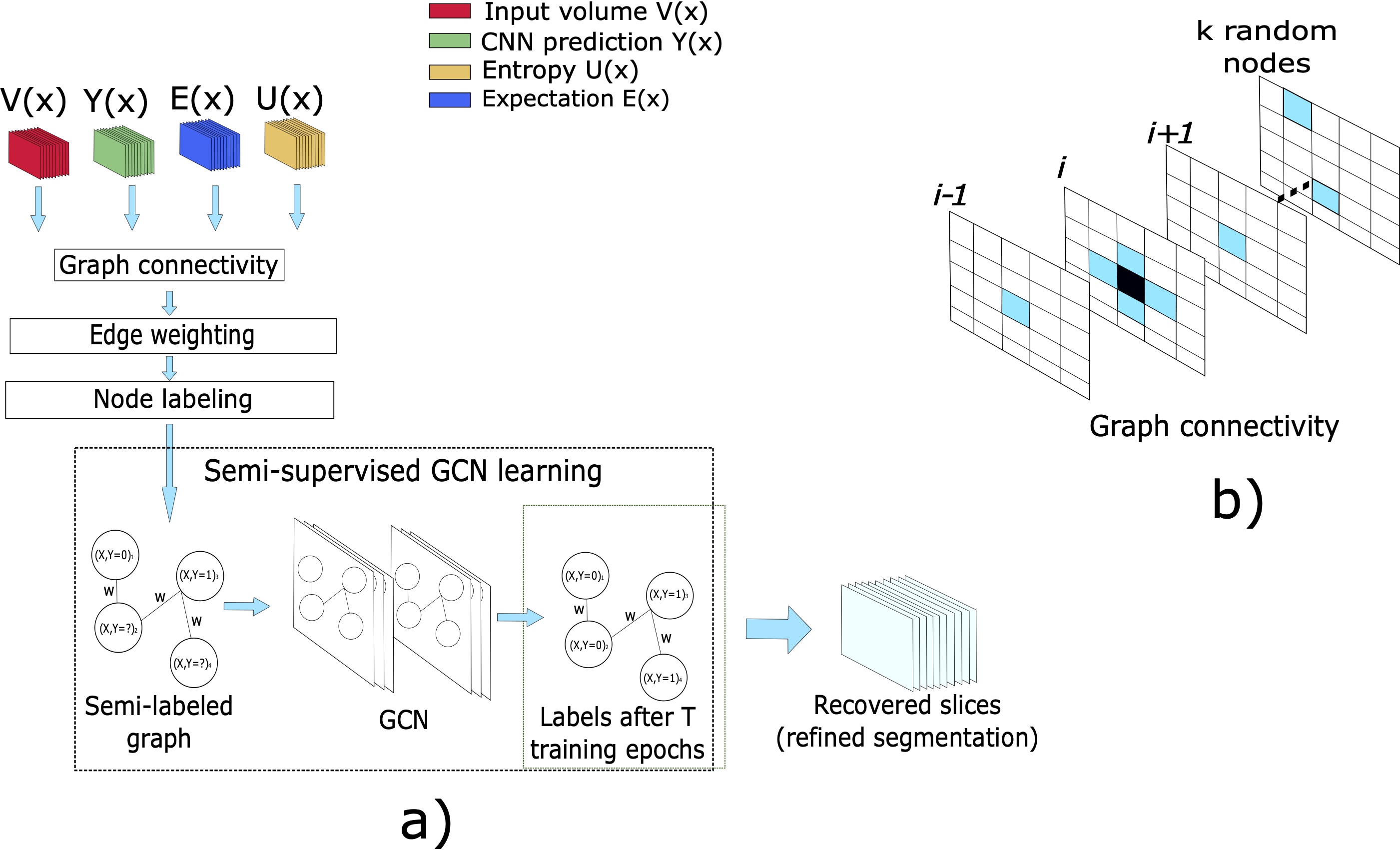}    
\end{center}
\caption{a) The GCN refinement strategy. We construct a semi-labeled graph representation based on the uncertainty analysis of the CNN.  Then, a GCN is trained to refine the segmentation. b) Connectivity. The black square is connected to six perpendicular neighbors and with $k=16$ random voxels} \label{fig:gcn}
\end{figure}

Given a graph $\mathcal{G}$ representing our 3D volumetric data, at the inference time, we aim to obtain a refined segmentation $Y^*$ as the results of our GCN model $\Gamma$,
\begin{equation}
Y^* = \Gamma(\mathcal{G}(S);  \phi),
\end{equation}
where the graph $\mathcal{G}$ is constructed from the set of volumes $S=\{\mathbb{E}, \mathbb{U}, V, Y\}$  (see section \ref{subsec:unc_analysis} and Fig. \ref{fig:gcn}) and $\phi$ represents the GCN's parameters.

Since most of the voxels in the volume are irrelevant for the refinement process and given that graphs are not restricted to the rectangular structured representation of data, we define an ROI tailored to our target anatomy. We define our working region as $\hbox{ROI}(x) = \hbox{dilation}(U_b(x)) \cup \mathbb{E}_b(x)$ with $\mathbb{E}_b$ the expectation binarized by a threshold of $0.5$. Since the entropy is usually high in boundary regions, including the dilated $U_b$ ensures that the ROI is bigger enough to contain the organ. Also, this allows us to include high confidence background predictions ($Y = 0$) for training the GCN. Including the expectation in the ROI give us high confidence foreground predictions for training the model. This ROI reduces the number of nodes of the graph and, in consequence, the memory requirements. 
The voxels $x \in \hbox{ROI}$ define the nodes for $\mathcal{G}$. Each node is represented by a feature vector containing intensity $V(x)$, expectation $\mathbb{E}(x)$, and  entropy  $\mathbb{U}(x)$. Finally, we labeled each node in the graph according to its uncertainty level using the next rule:
\begin{equation}
  l(x) =
  \begin{cases}
        Y(x) & \text{if $U_b(x)=0$} \\
  	    \text{unlabeled} & \text{if $U_b(x)=1$}
  \end{cases}
\end{equation}

\subsubsection{Edges and Weighting} \label{subsubsec:edges_weighting}

The most straightforward connectivity option is to consider the connectivity with adjacent voxels (n6 or n26 adjacent voxel neighborhood). However, this simple nearest neighborhood scheme may not be adequate for our problem for two reasons; First, with this scheme, every single voxel is connected with its local neighborhood but lacks global information. Considering the original volumetric representation of the data, this means that the main source of information is coming only from adjacent voxels, while information from the global context (not-adjacent voxels) is mainly ignored.

Second, voxels with high uncertainty tend to shape contiguous clusters. Because of this, a voxel with high uncertainty will be most probably surrounded by other high uncertainty points, reducing the connection with the low uncertainty points. Hence, a simple n6 or n26 connectivity might limit the propagation of information from the confidence to the uncertain regions. 

A  fully-connected graph can take advantage of the relationships between local and long-range connections, and propagate information from both certain and uncertain regions during training and inference time. The main disadvantage of a fully-connected graph for a GCN model is the prohibitive memory requirements. 

In our work, we evaluate an intermediate solution. For a particular node (or voxel) $x$, we create connections with its six perpendicular immediate neighbors in the volume coordinate system (N6) to consider local information. Additionally, we randomly select $k$ additional nodes in the graph and create a connection between these random elements and $x$. This defines a sparse representation that considers local and long-range connections between high and low uncertainty elements. The $k$ random nodes can be taken from any part inside the ROI used to define the nodes of the graph. We empirically found that $k=16$ offers a balance in performance and graph size, and kept this value during our experiments.

To define the weights for the edges, we use a function based on Gaussian kernels considering the intensity $V (x)$ and the 3-D position $x \in \mathbb{R}^3$ associated with the node:
\begin{equation} \label{eq:w1}
\begin{split}
w_1(x_i, x_j) = \lambda \hbox{div}(x_i, x_j) + \beta[\exp(-{\frac{||V (x) - V (x_j)||^2}{2\sigma_1} }) + \exp(-{ \frac{||x_i - x_j||^2}{2\sigma_2} })]
\end{split}
\end{equation}
where $\lambda$ and $\beta$ are balancing factors, $\hbox{div}(\cdot)$ is given by the diversity between the nodes \citep{bib:Zhouz17}, defined as $ \hbox{div} (x_i, x_j) = \sum_{c=1} ^M (P ^c(x_i) - P ^c(x_j))\log{\frac{P ^c(x_i) }{P ^c(x_j)}}$ with $M = 2$, $P^1(x_i) =\mathbb{E}(x_i)$ and $P ^2(x_i) = 1 - \mathbb{E}(x_i)$ for our binary case. We opt for an additive weighting, instead of a multiplicative one, because the GCN can take advantage of connections with both similar and dissimilar nodes (in intensity and space) in the learning process, and using a multiplicative weighting could completely cut these connections. We found out that the diversity can indirectly bring information about the similarity of two nodes, in terms of class probability.

Since the diversity does not have an upper bound, it is possible to apply a non-linear transformation in order to normalize its value to the range $(0, 1]$, leading to the following version of the diversity: 
\begin{equation} \label{eq:norm_div}
\hbox{norm\_div}(x_i, x_j) = 1 - 2^{-\hbox{div}(x_i, x_j)}.
\end{equation}

We can integrate this into eq.(\ref{eq:w1}) replacing the regular diversity:
\begin{equation} \label{eq:w2}
\begin{split}
w_2(x_i, x_j) = \lambda \hbox{norm\_div}(x_i, x_j) + \beta[\exp(-{\frac{||V (x) - V (x_j)||^2}{2\sigma_1} }) + \exp(-{ \frac{||x_i - x_j||^2}{2\sigma_2} })].
\end{split}
\end{equation}

Similarly, if we only take the exponential part of eq.(\ref{eq:norm_div}), we get a similarity metric between the expectation of the nodes $x_i$ and $x_j$. We use this version of the diversity as a third variation of our weighting function, leading to the following expression:
\begin{equation} \label{eq:w3}
\begin{split}
w_3(x_i, x_j) = \lambda \hbox{inv\_div}(x_i, x_j) + \beta[\exp(-{\frac{||V (x) - V (x_j)||^2}{2\sigma_1} }) + \exp(-{ \frac{||x_i - x_j||^2}{2\sigma_2} })],
\end{split}
\end{equation}
where the function $\hbox{inv\_div}(x_i, x_j)$ is given by
\begin{equation} \label{eq:inv_div}
\hbox{inv\_div}(x_i, x_j) = 2^{-\hbox{div}(x_i, x_j)}.
\end{equation}
It is worth mentioning that if we set $\lambda = 0$ and keep the same value of $\beta$  all the weighting functions become the same. 

In addition to the uncertainty threshold analysis presented in \cite{bib:soberanis20}, we extend this analysis to the different variations of the weighting functions introduced, exploring their effects in the refinement scheme. The corresponding analysis is presented in section \ref{sec:edge_weighting}. 

\subsubsection{Semi-Supervised GCN Learning}

At this point, we have reformulated the refinement task as a semi-supervised graph learning tasks. As mentioned previously, we use the proposal from \cite{bib:Kipf17} to train the GCN in a semi-supervised way. A convolutional $H$ layer is defined as:

\begin{equation}\label{eq:propagation}
    H = \hat{A}XW, \hbox{ with } \hat{A} = \tilde{D}^{-\frac{1}{2}}\tilde{A}\tilde{D}^{-\frac{1}{2}}
\end{equation}
The variable $X \in \mathbb{R} ^{N \times K}$ represents the input feature matrix of the layer with $N$ the number of nodes and $K$ the number of input features per node. $W \in \mathbb{R} ^{K \times K'}$ is the weight matrix of the current layer with $K'$ the number of output features per node. $\hat{A}$ follows the renormalization proposed by \cite{bib:Kipf17} with $\tilde{A} = A + I_N$, $A$ the adjacency matrix of $\mathcal{G}$, and the diagonal matrix $\tilde{D} = \Sigma_{\hbox{cols}} \tilde{A}$ that sums across the columns of $\tilde{A}$. In general, we keep the same GCN architecture, but employing a sigmoid activation function at the output layer, for binary classification. Representing our graph $\mathcal{G}$ by its adjacency matrix $A$ and feature matrix $X$, this leads to the following GCN model:
\begin{equation}
    Y^* = \Gamma(X, A; W^0, W^1) = \hbox{sigmoid}(\hat{A}\hbox{ }\hbox{ReLU}(\hat{A}XW^0)W^1)
\end{equation}

Finally, it is worth to mention that, even if we validated our proposal using this GCN learning architecture, given the modular nature of our method, it is possible to employ a different semi-supervised graph learning strategy.

\section{Experiments and Results}
    We validate our method refining the output of a 2D CNN in the tasks of pancreas and spleen segmentation.  We compare this approach with the refinement obtained from a conditional random field method \citep{bib:Krahenbuhl11}. Then, we evaluate the effects of variations in the graph-definition parameters, performing a sensitivity analysis.  All the processes run on an NVidia Titan Xp. We make our code publicly available for reproducibility purposes\footnote[1]{https://github.com/rodsom22/gcn\_refinement}. 

\subsection{Datasets}
We tested our framework using two CT datasets for pancreas, and spleen segmentation. For the pancreas segmentation problem, we used the NIH pancreas dataset\footnote[2]{https://wiki.cancerimagingarchive.net/display/Public/Pancreas-CT} \citep{bib:Roth16,bib:Roth15,bib:Clark13}. We randomly selected 45 volumes of the NIH dataset for training the CNN model and reserved 20 volumes for testing the uncertainty-based GCN refinement. 
For spleen, we employed the spleen segmentation task of the medical segmentation decathlon \citep{bib:simpson19} (MSD-spleen\footnote[3]{http://medicaldecathlon.com/}). For this problem, we trained the  CNN on 26 volumes and reserved nine volumes to test our framework. The MSD-spleen dataset contains more than one foreground label in the segmentation mask. We unified the non-background labels of the  MSD-spleen dataset into a single foreground class since we evaluate our method for refining a binary segmentation model. 

\subsection{Implementation Details}
\subsubsection{CNN Baseline Model}
We chose a 2D U-Net to be our CNN model \citep{bib:Ronneberger15}. We included dropout layers at the end of every convolutional block of the U-Net, as indicated by the MCDO method. The U-Net was trained considering a binary segmentation problem. Since we are employing a 2D model, we trained the models using axial slices. The model was trained with the dice loss function, using the Adam optimizer. We train the model for around 200 epochs keeping the overall best performing model during the entire training procedure. At inference time, we predicted every slice separately and then we stacked all the predictions together to obtain a volumetric segmentation (a similar strategy was used to perform the uncertainty analysis). As a post-processing step, we compute the largest connected component in the prediction, to reduce the number of false positives. At this point, it is worth mentioning that the U-Net was used only for testing purposes and different architectures can be used instead. This is mainly because our refinement method uses the model-independent MCDO analysis. 

\subsubsection{Uncertainty Analysis and GCN Implementation Details} 
We utilized MCDO to compute the expectation and entropy using a dropout rate of $0.3$ and a total of $T=20$ stochastic passes. To obtain volumetric uncertainty from a 2D model, we performed the uncertainty analysis on every individual slice of the input volume and then we stacked all the results together to obtain the volumetric expectation and entropy.  We tested different values for the uncertainty threshold $\tau$ (see section \ref{subsec:refinement-experimetns}).

The GCN model is a two-layered network with 32 feature maps in the hidden layer and a single output neuron for binary node-classification.  The graphical network is trained for 200 epochs with a learning rate of $1e-2$, binary entropy loss, and the Adam optimizer. We kept these same settings for the refinement of both segmentation tasks. After the refinement process, we can replace only the uncertain voxels with the GCN prediction, or we can replace the entire CNN prediction with the GCN output. We use the second approach since we found it producing better results. 

\subsubsection{Statistical Significance Test}

Given that with small sets, the statistical significance tests can fail \citep{bib:Szucs17,bib:biau2008}, we generate the dice score in slice-wise to increase the sample size (up with 278-1700 slices for the spleen, and pancreas respectively).  Then we have run the non-parametric statistical significance test, namely Kolmogorov–Smirnov test. We perform a statistical significance analysis between the results of the GCN refinement and the initial results of the CNN. A single start (*) indicates a p-value $<0.05$, while a double-start (**) indicates a p-value $< 0.01$ with respect to the original CNN prediction. 

\subsection{Comparison with State of the Art and Baseline CNN}

We applied our refinement method independently on every individual sample from the 20 NIH  and 9 MSD-spleen testing volumes. The edge weighting function for our refinement method is given by eq. \ref{eq:w1} with $\lambda=0.5$ and $\beta=1$. Since CRF is a common refinement strategy, we use the publicly available implementation of the method presented in \cite{bib:Krahenbuhl11} to refine the CNN prediction. This CRF method assumes dense connectivity. Similar to \cite{bib:Krahenbuhl11}, we set one unary and two pairwise potentials. We use the prediction of the CNN as the unary potential.  The first pairwise potential is composed of the position of the voxel in the 3D volume. The second pairwise potential is a combination of intensity and position of the voxels.  For the CRF refinement, we considered the same ROI used by the GCN. 
\begin{figure}[t]
\begin{center}
\includegraphics[width=0.82\textwidth]{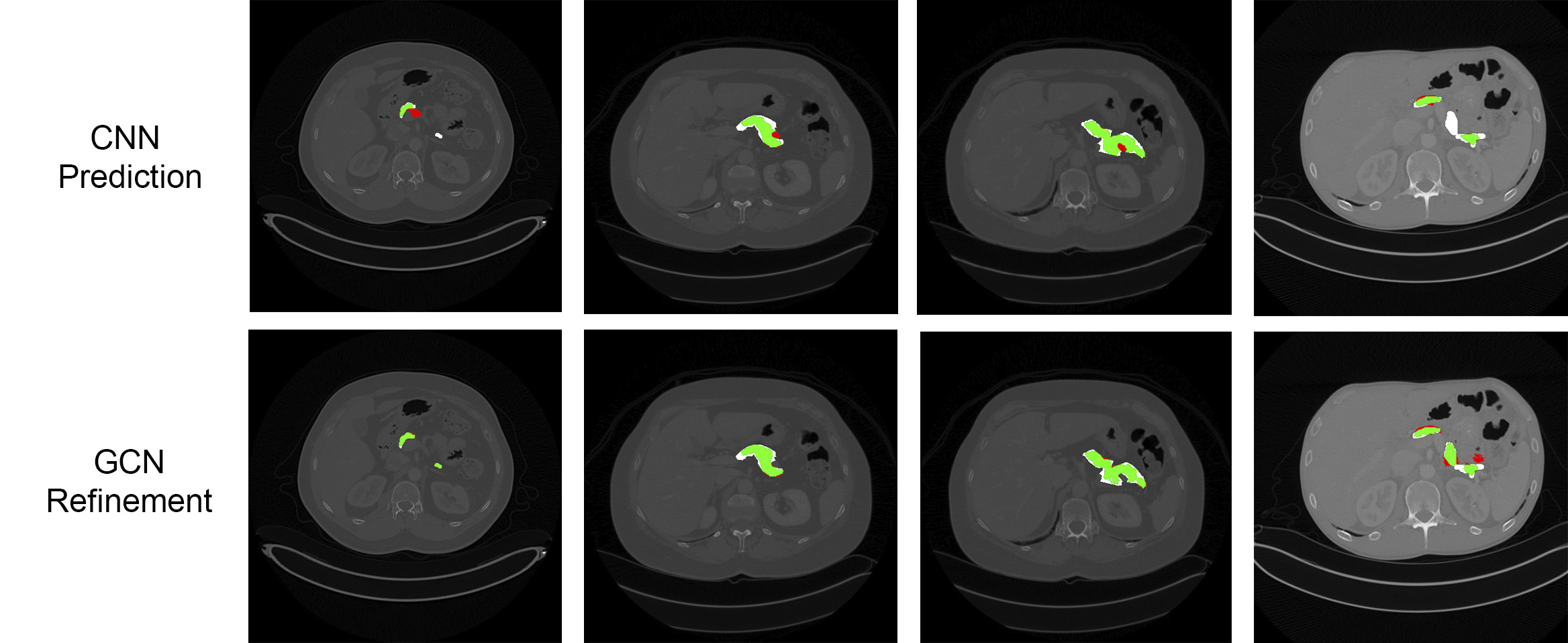}
\end{center}
\caption{Comparison of the CNN prediction and its corresponding GCN refinement for pancreas segmentation. Green colors indicate true positives (TP), red indicates false positives (FP), and white false negative (FN) regions. From left to right: the first column shows an FP region removed and an FN region recovered after the refinement.  The second and third columns show FP regions removed. The fourth column shows an FN region recovered but also a new FP region generated.} \label{fig:results_pancreas}
\end{figure}
\begin{table}[h]
			\caption{Average dice score performance (\%) of the GCN refinement compared with the CNN prediction and a CRF-based refinement of the CNN prediction. Results for pancreas and spleen are presented. Statistical significance is indicated by (*) for a p-value $<0.05$, and (**) for a p-value $< 0.01$ with respect to the CNN prediction.} 
			\label{tab:comparison}
			\centering
			\begin{tabular}{l | c | c | c }
			\hline 
			Task  & CNN & CRF & GCN    \\ 
			      & 2D U-Net & refinement & Refinement (ours)  \\ 
			\hline
			Pancreas & $76.89 \pm 6.6$ & $77.20 \pm 6.5$ & $\mathbf{77.81 \pm 6.3}$* \\ 
			\hline
			\hline
			Spleen & $93.17 \pm 2.5$ & $93.40 \pm 2.6$ & $\mathbf{95.07 \pm 1.3}$** \\  
			\hline
			\end{tabular}
\end{table}
\begin{figure}[h]
\begin{center}
\includegraphics[width=0.82\textwidth]{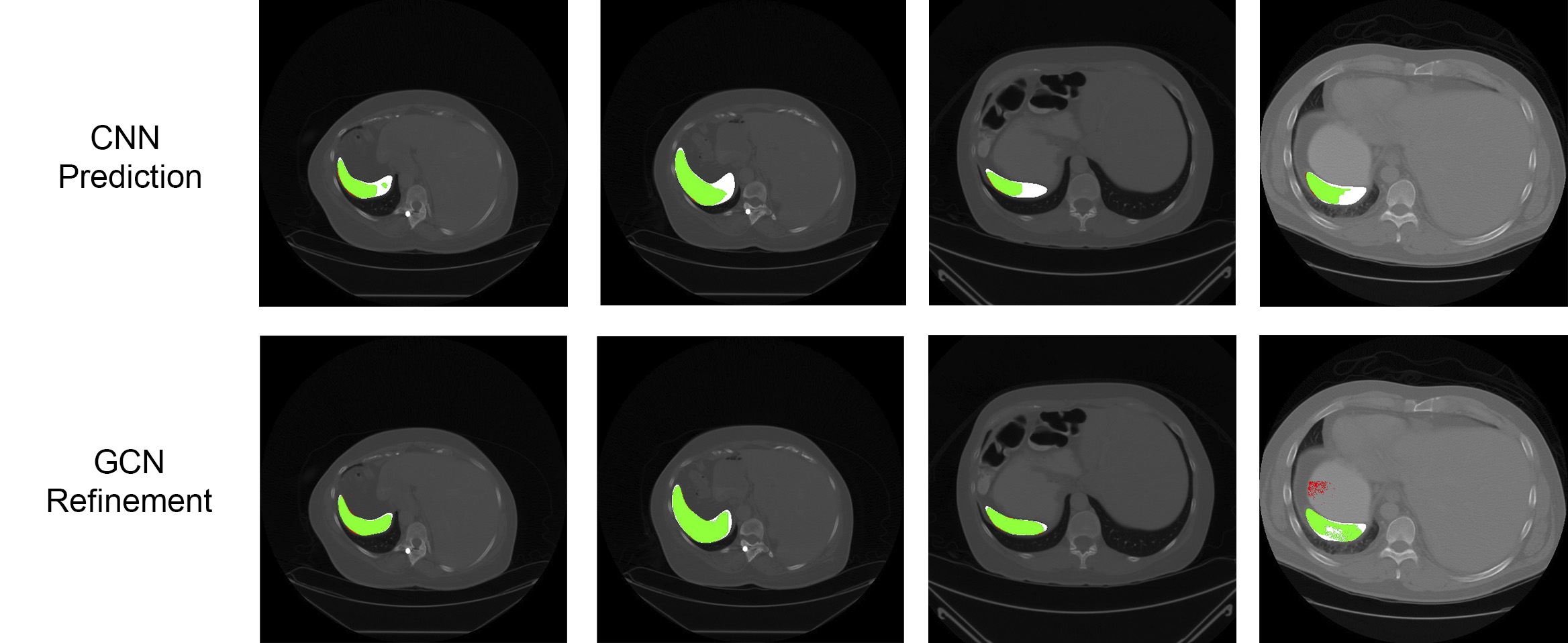}
\end{center}
\caption{Comparison of the CNN prediction and its corresponding GCN refinement for spleen segmentation. Green colors indicate true positives (TP), red indicates false positives (FP), and white false negative (FN) regions. From left to right: the first, second, and third columns show FN regions recovered. The fourth column shows an FN region recovered but also a new FP region generated.} \label{fig:results_spleen}
\end{figure}

Results are presented in Table \ref{tab:comparison}. The GCN-based refinement outperforms the base CNN model and the CRF refinement by around 1\% and 0.6\% respectively in the pancreas segmentation task. For spleen segmentation, our GCN refinement presented an increase in the dice score of 2\% with respect to the base CNN, and 1.7\% with respect to the CRF refinement. Figs. \ref{fig:results_pancreas} and \ref{fig:results_spleen} show visual examples of the GCN refinement compared with the base CNN prediction.

\subsection{Sensitivity Analysis}\label{sec:sen_analisys}
We performed an analysis of the performance of our proposal under different variations of their main components and hyper-parameters. Our analysis evaluate the scenario when the base CNN was trained with limited data (sec. \ref{subsec:refinement-numsamples}). We also explore how the choice of the uncertainty threshold can affect the performance (sec. \ref{subsec:refinement-experimetns}). A node connectivity considering the 26 surrounding neighborhood is compared with the n6 plus 16 long-range random connection we employed (sec. \ref{sec:node_conn}). Finally, a deep analysis on different variations of the weighting function is presented in sections \ref{sec:edge_weighting}, \ref{sec:div_simi}, \ref{sec:norm_div}, \ref{sec:all_simi}, and \ref{sec:div_only}. 

\subsubsection{Influence of the Number of Training Samples} \label{subsec:refinement-numsamples}
We evaluate the performance of the GCN refinement when the base CNN is trained with a small number of samples. For this, we randomly selected 10 out of the 45 training samples of the NIH dataset. For spleen, we selected nine. Results are presented in Table \ref{tab:reduce_training}.
\begin{table}[h]
			\caption{Average dice score performance (\%) of the GCN refinement compared with the CNN prediction. The CNN model was trained with 10 samples for the pancreas and 9 for the spleen. Statistical significance is indicated by (*) for a p-value $<0.05$, and (**) for a p-value $< 0.01$ with respect to the CNN prediction.} 
			\label{tab:reduce_training}
			\centering
			\begin{tabular}{l | c | c | c }
			\hline 
			Task  & CNN & CRF & GCN    \\ 
			      & 2D U-Net & refinement & Refinement (ours)  \\ 
			\hline
			Pancreas-10 & $52.14 \pm 22.6$ & $52.20 \pm 22.6$ & $\mathbf{54.55 \pm 22.2}$* \\ 
			\hline
			Spleen-9 & $78.89 \pm 28.4$ & $78.80 \pm 28.4$ & $\mathbf{81.15 \pm 28.9}$** \\ 
			\hline
			\end{tabular}
\end{table}

Note the increment in the standard deviation of all the models. A reason for this can be that the CNN does not generalize adequately to the testing set, due to the small number of training examples. Similar to the previous results, the increment in the dice score for the GCN refinement is about 2.4\% with respect to the CNN base model for the pancreas, and improvement of 2.3\% for spleen, compared with the base CNN.

\subsubsection{Influence of Uncertainty Threshold} \label{subsec:refinement-experimetns}
In our experiments, we evaluate the influence of different values for $\tau$. We tested the method with values of $\tau \in \{0.001, 0.3, 0.5, 0.8, 0.999\}$. In this way, we covered a wide range of conditions that define a voxel as ``uncertain''. After training the GCN, we replaced all the CNN predictions with the GCN output. Table \ref{tab:cnn_vs_gcnref} compares the CNN output with the GCN refinement at different values of $\tau$ for the tasks of the pancreas and spleen segmentation. 

\begin{table}
			\caption{Average dice score performance (\%) of the GCN refinement at different uncertainty thresholds $\tau$. Pancreas-10 and Spleen-9 indicate the models trained with 10 and nine samples, respectively.} 
			\label{tab:cnn_vs_gcnref}
			\centering
			\begin{tabular}{l | c | c | c | c | c }
			\hline 
			Task  &  GCN & GCN & GCN & GCN & GCN  \\ 
			      & $\tau = 1e-3$ & $\tau = 0.3$ & $\tau = 0.5$ & $\tau = 0.8 $ & $\tau = 0.999$  \\ 
			\hline
			Pancreas & $77.71 \pm 6.3$ & $77.79 \pm 6.4$ & $77.77 \pm 6.3$ & $77.81 \pm 6.3$ & $77.79 \pm 6.3$ \\ 
			\hline
			Pancreas-10 & $54.55 \pm 22.1$ & $54.32 \pm 22.1$ & $54.15 \pm 22.2$ & $53.91 \pm 22.4$ & $53.14 \pm 22.9$ \\ 
			\hline
			\hline
			Spleen & $95.01 \pm 1.5$ & $94.92 \pm 1.4$ & $94.98 \pm 1.4$ & $94.97 \pm 1.4$ & $95.07 \pm 1.3$ \\  
			\hline
			Spleen-9 & $80.91 \pm 28.8$ & $80.94 \pm 28.9$ & $80.94 \pm 28.8$ & $80.98 \pm 28.9$ &  $81.15 \pm 28.9$\\ 
			\hline
			\end{tabular}
\end{table}

The parameter $\tau$ controls the minimum requirement to consider a voxel as uncertain. Lower values lead to a higher number of uncertain elements. This has a direct relationship with the number of high certainty nodes in the graph representation, and hence, in the number of training examples for the GCN. This also influences the quality of the training voxels for the GCN, since a high threshold relaxes the amount of uncertainty necessary to rely on the prediction of the CNN.

However, from the results of Table \ref{tab:cnn_vs_gcnref}, except for pancreas-10 and spleen-9, there is no significant impact on the choice of this parameter. One reason can be that there is a clear separation between high and low uncertainty points. Therefore, changing $\tau$ may add (remove) a few number of nodes that are insignificant for the learning process of the GCN. 

For the pancreas-10 model, we notice a progressive decrease in the dice score. Since this model uses fewer training examples, it is expected to have low confidence in their predictions (in contrast with the model trained with 45 volumes). In this scenario, a higher uncertainty threshold increases the chance to include high-uncertainty nodes as ground truth for training the GCN. A lower $\tau$ includes fewer points but with higher confidence. This appears to be beneficial in the pancreas segmentation model trained with fewer examples. 
\begin{figure}[h]
\begin{center}
\includegraphics[width=0.48\textwidth]{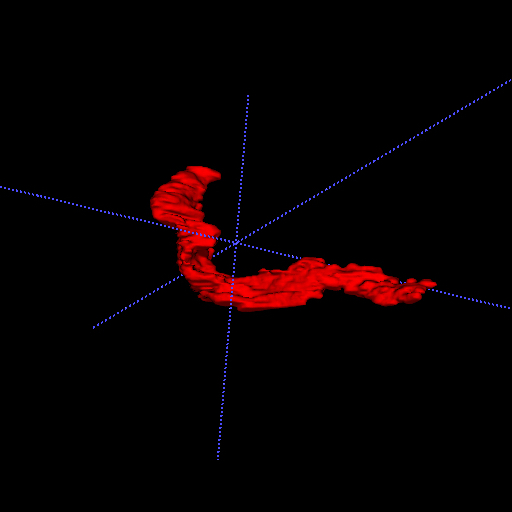}
\includegraphics[width=0.48\textwidth]{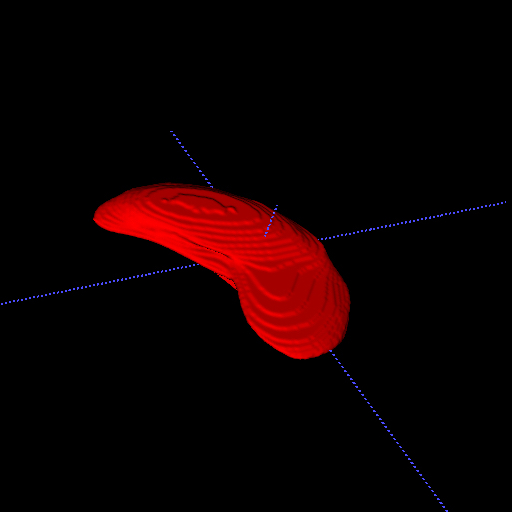}
\end{center}
\caption{An example of the 3D view of the pancreas and spleen. The shape of the pancreas brings more inter-patient and inter-slice variability. The shape of the spleen presents better consistency between patients and slices, leading to better performance for the CNNs. Views obtained with ITK-SNAP \citep{bib:Yushkevich2006}.} \label{fig:pancreas_spleen}
\end{figure}

The opposite occurs with spleen-9, where higher $\tau$ are beneficial. This might indicate a dependency on the characteristics of the anatomies since the pancreas presents more inter-patient and inter-slice variability (see Fig. \ref{fig:pancreas_spleen}).

In general, our results suggest that $\tau$ parameter should be selected based on the target anatomy. Further, $\tau$ appears to have more influence in conditions of high uncertainty, e.g. when the model is trained with fewer examples. In the cases where $\tau$ has no significant impact, intermediate values are preferred, since they lead to a lower number of nodes, and in consequence to lower memory requirements. 

\subsubsection{Node Connectivity}\label{sec:node_conn}

We compare the refinement performance of our method when a classical n26 neighborhood is employed. We repeated the experiments for different uncertainty thresholds, and the CNN model trained with different numbers of examples on the pancreas and spleen datasets. Results are presented in Tables \ref{tab:n6n26p} and \ref{tab:n6n26s} for each organ, respectively. In all tables, the weighting employed is eq. \ref{eq:w1} with $\lambda=0.5$ and $\beta=1$.

\begin{table}[h]
			\caption{Average dice score performance (\%) of the GCN refinement at different uncertainty thresholds $\tau$ for the \textbf{pancreas} segmentation problem. The table compares the results for the 6-surrounding + 16 random connectivity (ours) vs. a 26 surrounding connectivity (n26). cnn10 indicates results obtained with a CNN trained on ten samples.} 
			\label{tab:n6n26p}
			\centering
			\begin{tabular}{l | c | c | c | c | c }
			\hline 
			Connectivity  &  GCN & GCN & GCN & GCN & GCN  \\ 
			              & $\tau = 1e-3$ & $\tau = 0.3$ & $\tau = 0.5$ & $\tau = 0.8 $ & $\tau = 0.999$  \\ 
			\hline
			Ours & $77.71 \pm 6.3$ & $77.79 \pm 6.4$ & $77.77 \pm 6.3$ & $77.81 \pm 6.3$ & $77.79 \pm 6.3$ \\ 
			\hline
			n26& $ 76.93 \pm 6.1$ & $77.16 \pm 5.8$ & $77.18 \pm 5.8$ & $77.18 \pm 5.8$ & $77.4 \pm 5.9$ \\ 
			\hline
			\hline
			Ours-cnn10& $54.55 \pm 22.1$ & $54.32 \pm 22.1$ & $54.15 \pm 22.2$ & $53.91 \pm 22.4$ & $53.14 \pm 22.9$ \\ 
			\hline
			n26-cnn10& $52.50 \pm 22.1$ & $52.66 \pm 22.4$ & $52.56 \pm 22.4$ & $52.53 \pm 22.5$ &  $52.43 \pm 22.9$\\ 
			\hline
			\end{tabular}
\end{table}
\begin{table}[h]
			\caption{Average dice score performance (\%) of the GCN refinement at different uncertainty thresholds $\tau$ for the \textbf{spleen} segmentation problem. The table compares the results for the 6-surrounding + 16 random connectivity (ours) vs. a 26 surrounding connectivity (n26). cnn9 indicates results obtained with a CNN trained on nine samples.} 
			\label{tab:n6n26s}
			\centering
			\begin{tabular}{l | c | c | c | c | c }
			\hline 
			Connectivity  &  GCN & GCN & GCN & GCN & GCN  \\ 
			      & $\tau = 1e-3$ & $\tau = 0.3$ & $\tau = 0.5$ & $\tau = 0.8 $ & $\tau = 0.999$  \\ 
			\hline
			Ours & $95.01 \pm 1.5$ & $94.92 \pm 1.4$ & $94.98 \pm 1.4$ & $94.97 \pm 1.4$ & $95.07 \pm 1.3$ \\ 
			\hline
			n26 & $94.26 \pm 1.6$ & $94.11 \pm 1.8$ & $95.09 \pm 1.9$ & $94.14 \pm 1.8$ & $94.43 \pm 1.8$ \\ 
			\hline
			\hline
			Ours-cnn9 & $80.91 \pm 28.8$ & $80.94 \pm 28.9$ & $80.94 \pm 28.8$ & $80.98 \pm 28.9$ &  $81.15 \pm 28.9$\\  
			\hline
			n26-cnn9 & $79.63 \pm 28.6$ & $79.65 \pm 28.5$ & $79.68 \pm 28.5$ & $79.90 \pm 28.5$ &  $80.38 \pm 28.7$\\ 
			\hline
			\end{tabular}
\end{table}

The results show a better refinement when including the random long-range connections, especially when working with the CNN trained with limited data, for both tasks.  It is worth to mention that when comparing with the original CNN output presented in Tables \ref{tab:comparison} and \ref{tab:reduce_training}, our refinement method using the n26 connectivity still having a slight improvement over the original CNN prediction.

\subsubsection{Edge Weighting} \label{sec:edge_weighting}

Finally, we analyze the influence of different variations of the components of our weighting function. We perform our refinement strategy using the weighting functions given by equations \ref{eq:w1}, \ref{eq:w2}, and \ref{eq:w3}, presented in section \ref{subsubsec:edges_weighting}. In all the experiments of this section, we keep a fixed value of $\tau=0.5$. We will use the notation $w_{i,(\lambda,\beta)}$, $i \in \{1, 2, 3\}$ to indicate the parameters employed by the corresponding function, for example, the notation $w_{1,(0.5,1)}$ holds for weighting function $w_1$ with $\lambda=0.5$ and $\beta=1$. We explore the situations when: a) either only diversity or only the Gaussian kernels are employed (sec. \ref{sec:div_simi}); b) using a diversity normalized to the range [0, 1] (sec. \ref{sec:norm_div}); c) using similarity in expectation given by the inverse of the diversity  (sec. \ref{sec:all_simi}); and d) we discuss deep insights regarding the three different variations of the diversity employed in the previous sections (sec. \ref{sec:div_only}).

\subsubsection{Weighting: Diversity and Gaussian Similarity Kernels} \label{sec:div_simi}

We can divide the weighting function into two metrics, diversity and Gaussian similarity kernels (in intensity and position). In this experiment, we keep one of these components at a time. This is done by setting the values of $\lambda$ and $\beta$ to zero, accordingly. The results are presented in Table \ref{tab:div_simi}. Note that $w_{1,(0.5,1)}$ corresponds to the weighting function used so far. 
\begin{table}[h]
			\caption{Average dice score performance (\%) of the GCN refinement at $\tau=0.5$. The table compares $w_1$ with different combinations of $\lambda$, $\beta$ parameters ($w_{1,(\lambda,\beta)}$). Statistical significance is indicated by (*) for a p-value $<0.05$, and (**) for a p-value $< 0.01$ with respect to the CNN prediction.} 
			\label{tab:div_simi}
			\centering
			\begin{tabular}{l | c | c | c | c }
			\hline 
			Task  &     CNN      & GCN             &  GCN            &  GCN  \\ 
			      &  2D U-Net    & $w_{1,(0.5,1)}$ & $w_{1,(0.5,0)}$ & $w_{1,(0,1)}$\\
			\hline
			Pancreas & $76.89 \pm 6.6$ & $77.77 \pm 6.3$* & $77.85 \pm 6.3$*  & $77.90 \pm 6.2$* \\ 
			\hline
			Pancreas-10 & $52.14 \pm 22.6$ & $54.15 \pm 22.2$ & $54.28 \pm 22.2$  & $53.03 \pm 23.0$ \\ 
			\hline
			\hline
			Spleen & $93.17 \pm 2.5$ & $94.98 \pm 1.4$** & $95.20 \pm 1.4$** & $94.74 \pm 1.8$** \\ 
			\hline
			Spleen-9 & $78.89 \pm 28.4$ & $80.94 \pm 28.8$** & $80.96 \pm 28.9$**  & $81.17 \pm 28.9$*\\ 
			\hline
			\end{tabular}
\end{table}

Different weighting functions and variations still outperform the initial CNN prediction. Now, we will focus on the small differences between these results. As we mentioned, our weighting scheme considers diversity together with intensity and position similarity. The two later Gaussian components are commonly used in the literature and follow the intuition that two components that are similar in intensity and close to each other are likely to belong to the same class. The diversity is an additional component that allows us to include the results from the MCDO analysis in the edge weighting. The diversity has a lower bound of zero but in contrast with the Gaussian kernels, in an ideal form, it does not have an upper bound. When two nodes have a similar expectation, the diversity between those elements will be small, and the weighting function will only rely on the Gaussian similarities. When the nodes have important differences in expectation (e.g 0 vs. 1), the unbounded nature of the diversity will ignore the much smaller contribution of the Gaussian kernels (the diversity can reach values of $30$, while each Gaussian kernel has a maximum at $1.0$), and bias the weight to the value of the diversity.

\paragraph{Pancreas models.} In a high-training-data scenario,  a Gaussian-kernel-only ($\lambda=0$) scheme appears to be good enough for the refinement strategy. In the low data-regime, the GCN appears to take more advantage of diversity. Our features employ intensity, expectation, and entropy (or uncertainty). Note that the uncertainty threshold affects the labeled nodes but does not affect the connectivity. A node can be connected to any certain or uncertain neighbor, and a Gaussian-kernel-only weighting is not aware of possible inconsistencies in the node features. 

\paragraph{Spleen models.} The spleen segmentation problem shows different behavior. The lower inter-patient variability of the spleen could be a reason (see Figs. \ref{fig:pancreas_spleen}).  In a high-training-data regime, it can bring a more stable and well-separated expectation between organ and background, making the diversity a more favorable weighting. In a low data regime, the Gaussian-kernel-only version appears to perform better. However, from Table 3,  a larger $\tau$ appears to benefit the low-data spleen problem. In fact, at $\tau=0.999$, the dice-score for $w_{1,(0.5,0)}$ and $w_{1,(0,1)}$ are 81.09, and 81.15, respectively, showing no significant difference. Note that $w_{1,(0,1)} = w_{2,(0,1)} = w_{3,(0,1)}$. 

\subsubsection{Normalizing the Diversity} \label{sec:norm_div}

One of the questions, we had, during this work whether the normalization of the diversity would have a positive/negative impact on the behavior of the GCN. In fact, $w_2$ presents a negative impact on the learning process. As we mentioned, the unbounded nature of the diversity makes $w_1$ to prefer connections with opposed expectations (and ignore the Gaussian similarity in those cases). On the other hand, the Gaussian similarity is only considered when the expectation of both nodes is basically the same, and even in those cases, the weight is small compared with the values of the diversity. In this sense, the GCN can learn mostly from examples that are different in expectation or (with a considerably lower contribution ) from examples similar in intensity and position. 

Using $w_2$, which normalizes the diversity into a value between 0 and 1, makes the contribution of the Gaussian kernels more representative. The normalized diversity of $w_2$ will no longer ignore the Gaussian similarity, and it will assign the highest weights to connections between nodes that are similar in intensity and position, but at the same time different in expectation, which is counter-intuitive. Table \ref{tab:norm_div} shows the drop in the performance using $w_2$ as weighting function. Given that the diversity is in range between 0 and 1, we use $\lambda = 1$ for $w_2$.
\begin{table}[h]
			\caption{Average dice score performance (\%) of the GCN refinement at $\tau=0.5$. The table shows the drop in the performance of $w_2$ compared with the U-Net prediction and $w_1$. Statistical significance is indicated by (*) for a p-value $<0.05$, and (**) for a p-value $< 0.01$ with respect to the CNN prediction.} 
			\label{tab:norm_div}
			\centering
			\begin{tabular}{l | c | c | c }
			\hline 
			Task  &     CNN      & GCN             &  GCN  \\ 
			      &  2D U-Net    & $w_{1,(0.5,1)}$ & $w_{2,(1,1)}$ \\
			\hline
			Pancreas & $76.89 \pm 6.6$ & $77.77 \pm 6.3$* & $63.27 \pm 9.9$**  \\ 
			\hline
			Pancreas-10 & $52.14 \pm 22.6$ & $54.15 \pm 22.2$ & $43.05 \pm 22.1$**  \\ 
			\hline
			\hline
			Spleen & $93.17 \pm 2.5$ & $94.98 \pm 1.4$** & $87.15 \pm 3.1$**  \\ 
			\hline
			Spleen-9 & $78.89 \pm 28.4$ & $80.94 \pm 28.8$** & $75.47 \pm 26.8$** \\ 
			\hline
			\end{tabular}
\end{table}

\subsubsection{All-Similarity Weighting} \label{sec:all_simi}

If we take only the exponential part of eq. \ref{eq:norm_div}, we will obtain values in the range [0,1] that gives high weights to similar expectations, presenting a better agreement with the Gaussian kernels. This weighting scheme is given by $w_3$. The results, compared with the U-Net and the initial $w_{1,(0.5,1)}$ weighting function are presented in Table \ref{tab:inv_div}. Results, again show an improvement for the GCN refinement, supporting our discussion about $w_2$. 
\begin{table}[t]
			\caption{Average dice score performance (\%) of the GCN refinement at $\tau=0.5$. The table compares the performance of $w_3$ with the U-Net prediction and $w_1$. Statistical significance is indicated by (*) for a p-value $<0.05$, and (**) for a p-value $< 0.01$ with respect to the CNN prediction.} 
			\label{tab:inv_div}
			\centering
			\begin{tabular}{l | c | c | c }
			\hline 
			Task  &     CNN      & GCN             &  GCN  \\ 
			      &  2D U-Net    & $w_{1,(0.5,1)}$ & $w_{3,(1,1)}$ \\
			\hline
			Pancreas & $76.89 \pm 6.6$ & $77.77 \pm 6.3$* & $78.19 \pm 6.1$*  \\ 
			\hline
			Pancreas-10 & $52.14 \pm 22.6$ & $54.15 \pm 22.20$ & $52.90 \pm 23.1$  \\ 
			\hline
			\hline
			Spleen & $93.17 \pm 2.5$ & $94.98 \pm 1.4$** & $94.81 \pm 1.9$**  \\ 
			\hline
			Spleen-9 & $78.89 \pm 28.4$ & $80.94 \pm 28.8$** & $81.08 \pm 28.9$** \\ 
			\hline
			\end{tabular}
\end{table}

\paragraph{Pancreas models.} Again, we can first focus on the results of the pancreas. The inverse of the diversity in combination with the Gaussian kernels appears to perform well with the pancreas model trained with a high number of examples. This can be because of the number of examples is enough to derive a good estimation of the expectation, in contrast with the low-data pancreas U-Net. 

\paragraph{Spleen models.} For the spleen problem, the differences appear to be not significant. In general, the results suggest that $w_3$ is beneficial for the full-data pancreas problem, and does work well with both spleen models. But might be sub-optimal for the low-data pancreas refinement task. A possible explanation is that $w_3$ will assign high weights to nodes with similar values, no matter if both have expectations of 1, 0, or 0.5. This last expectation value (0.5) represents a high uncertainty point, and it is expected to find this kind of point on an irregular organ, and with a model trained with a low number of examples, like Pancreas-10. In this sense, the inverse of the diversity in $w_3$ might also assign a high weight to connections between uncertainty nodes.

\subsubsection{Diversity-Only Weighting} \label{sec:div_only}

In this section, as final analysis for the weighting function, we compare $w_1$, $w_2$, and $w_3$ when $\beta=0$. The results are presented in Table \ref{tab:div_comp}. Two points are worth mentioning.  Comparing the results of $w_{2,(1,0)}$ in Table \ref{tab:div_comp} with the results of $w_{2,(1,1)}$ in Table \ref{tab:norm_div}, it is clear that the only-diversity version of $w_2$ has better performance, supporting the idea of the inconsistent connections of $w_{2,(1,1)}$. 
\begin{table}[h]
			\caption{Average dice score performance (\%) of the GCN refinement at $\tau=0.5$. The table compares the performance of the three variations of the diversity ($\beta=0$ for all functions). Statistical significance is indicated by (*) for a p-value $<0.05$, and (**) for a p-value $< 0.01$ with respect to the CNN prediction.} 
			\label{tab:div_comp}
			\centering
			\begin{tabular}{l | c | c | c | c }
			\hline 
			Task  &     CNN      & GCN             &  GCN            &  GCN  \\ 
			      &  2D U-Net    & $w_{1,(0.5,0)}$ & $w_{2,(1,0)}$ & $w_{3,(1,0)}$\\
			\hline
			Pancreas & $76.89 \pm 6.6$ & $77.85 \pm 6.3$* & $75.45 \pm 7.1$**  & $77.70 \pm 6.2$* \\ 
			\hline
			Pancreas-10 & $52.14 \pm 22.6$ & $54.28 \pm 22.2$ & $50.72 \pm 23.5$*  & $52.09 \pm 23.0$ \\ 
			\hline
			\hline
			Spleen & $93.17 \pm 2.5$ & $95.20 \pm 1.4$** & $94.25 \pm 2.1$** & $94.59 \pm 2.0$** \\ 
			\hline
			Spleen-9 & $78.89 \pm 28.4$ & $80.96 \pm 28.9$** & $79.84 \pm 28.5$  & $80.81 \pm 28.7$*\\ 
			\hline
			\end{tabular}
\end{table}

\begin{figure}
\begin{center}
\includegraphics[width=0.99\textwidth]{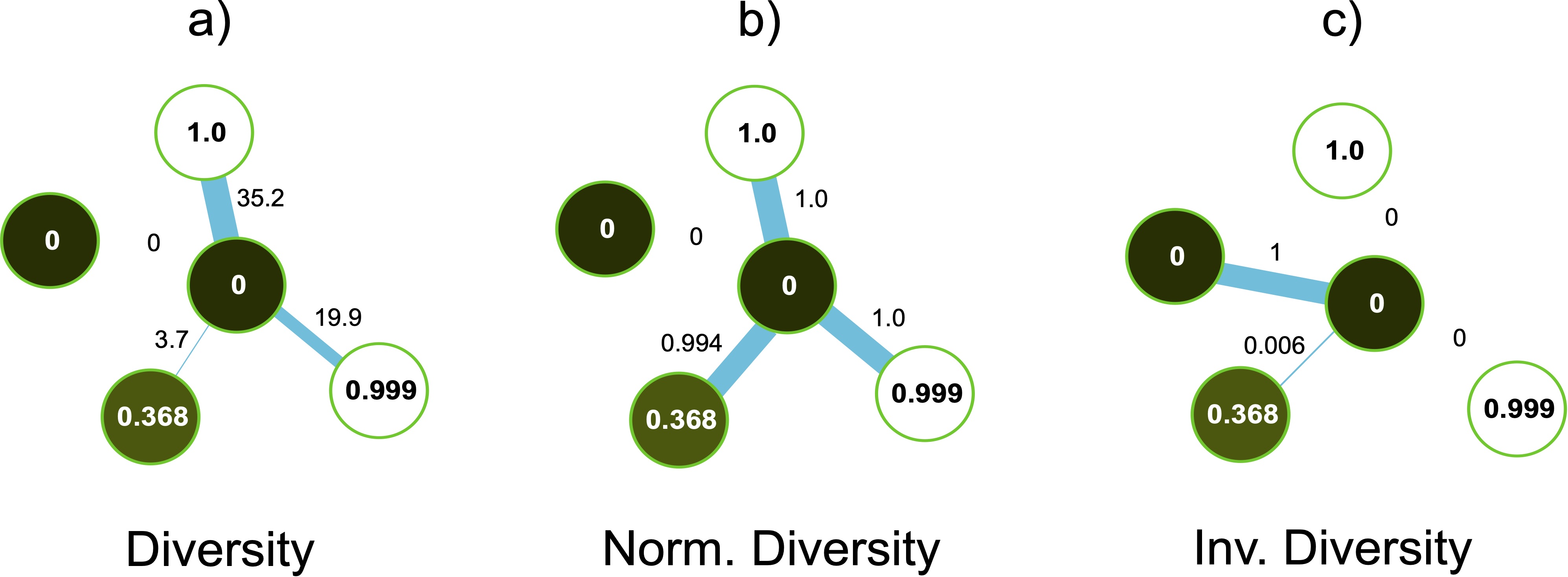}
\end{center}
\caption{A graphical view of the weights assigned by each variation of the diversity function (vanilla, normalized, and inverse).  The three figures show the same node structures, but weighted by: a)  $w_{1,(0.5,0)}$, b) $w_{2,(1,0)}$, and c) $w_{3,(1,0)}$. The value inside the node represents the expectation for that node.} \label{fig:div_weighting}
\end{figure}

\begin{figure}
\begin{center}
\includegraphics[width=0.95\textwidth]{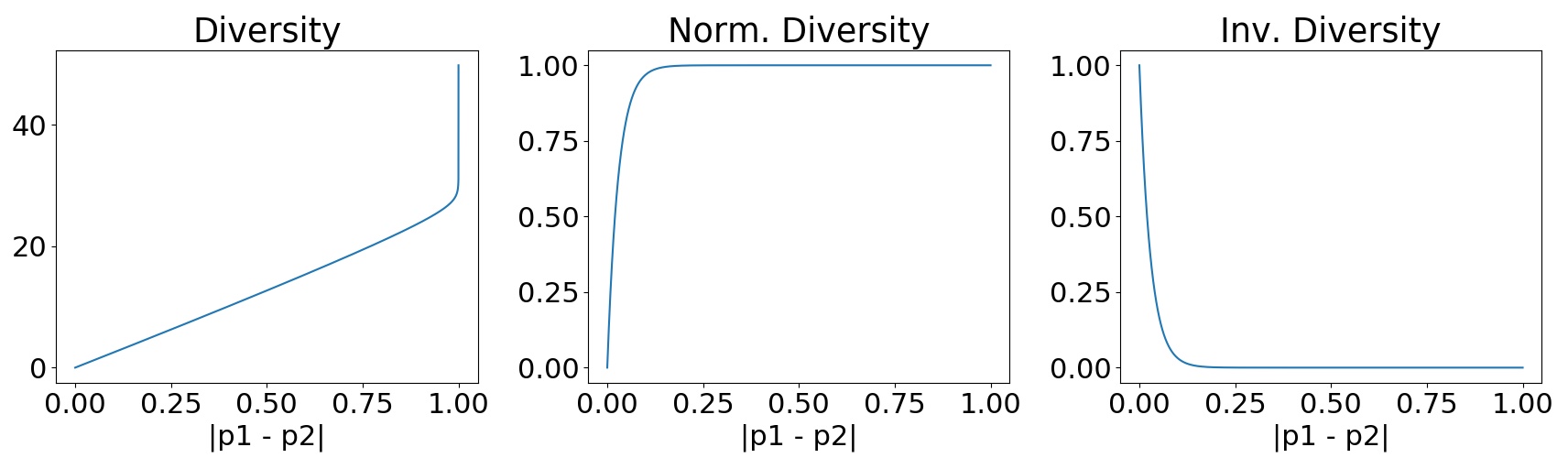}
\end{center}
\caption{Comparison of the weights assigned by the three variations of the diversity (vanilla, normalized, and inverse), represented as a function of the difference of the expectation $p_1=0$ and $p_2\in[0,1]$.} \label{fig:div_function}
\end{figure}

The second interesting fact comes when comparing the diversity of $w_1$ with the normalized diversity of $w_2$, both in Table \ref{tab:div_comp}. Even though these two functions are expected to behave similarly, we notice a difference in their performance. To understand the possible causes of these differences, we take a closer look at the assigned weights of these functions to a subset of nodes. Let's consider a central node with expectation close to zero, and four neighbors nodes with expectations of approximately 0.0, 1.0, 0.999, and 0.368, respectively (see Fig. \ref{fig:div_weighting}). As expected, $w_{1,(0.5,0)}$ will assign high values to opposed expectations (36.2 and 19.9 in Figure \ref{fig:div_weighting}.a), while the weight will be low for similar expectations, and close to zero to identical expectations (weights of 3.7 and 0 in Fig. \ref{fig:div_weighting}.a). However, for the same node structure,  $w_{2,(1,0)}$ will weight with a high value (close to 1) almost all the connections, except for the neighbors with the same expectation (see Fig, \ref{fig:div_weighting}.b). Even though the difference between the expectations values of 1 and 0.368 is not high enough, $w_{2,(1,0)}$  will give an importance that is similar to the nodes with completely opposed expectations, leading to an inconsistent weighting scheme. In contrast, $w_{3,(1,0)}$ (which acts as a kind of similarity in expectation) assigns weights close to zero to all the nodes that do not have the same expectation, even to the pair (0, 0.368), see Fig. \ref{fig:div_weighting}.c. This can suggest a better consistency for the inverse diversity, compared with diversity normalized to [0, 1].
We can also see these conclusions if we plot the values for these three variations of the diversity as a function of the difference between $p1= 0$ and $p2 \in [0, 1]$ (see Fig. \ref{fig:div_function}).  We can see how the diversity grows exponentially when the difference between $p1$ and $p2$ is close to 1 (Fig. \ref{fig:div_function} Diversity).  In a similar way, the inverse diversity assigns a weight of 1 to differences close to zero and decreases exponentially at the moment this difference starts to increment (Fig. \ref{fig:div_function} Inv. Diversity).  Finally, the normalized diversity starts assigning a weight of zero to equal expectations, however, the curve grows exponentially until reaching a weight of 1 when the difference in expectations still close to 0.2 (Fig. \ref{fig:div_function} Norm. Diversity), leading to the inconsistencies mentioned before. 

In general, a weighting function like $w_2$ appears to be not recommendable for the refinement with GCNs.  The inverse diversity of $w_3$ can be a better option, however, it is also possible that under certain conditions, this weighting could introduce noisy connections. For example, when the expectation has values close to 0.5 for both nodes, the function will assign high weights, however, this might not heavily contribute to the GCN given the fact that these nodes might have high uncertainty as well. This suggests that, under high uncertainty environments, it is better to weigh based on dissimilarity in expectation. On the other hand, the inverse of diversity can take advantage of well-separated expectations. 

\subsection{Hyper-parameters Search}

We performed an exhaustive search across the different weighting functions, and uncertainty thresholds, leading to the results presented in Table \ref{tab:update_dsc}.

As noted, the improvements are not significantly different from the initial weighting function $w_{1,(0.5,1)}$ presented in our initial work \citep{bib:soberanis20}, which suggests that this weighting function is good enough for the presented tasks. Nevertheless, a hyper-parameter search, e.g. choice of the weighting function, and $\tau$, might lead to  better performance. This mainly depends on the structure of interest and its characteristics, though. 

\begin{table}[h]
			\caption{Average dice score performance (\%) of the GCN refinement for the pancreas and spleen segmentation refinement. The initial results from \cite{bib:soberanis20} are compared with the best scores obtained after a $\tau$, $w_{i,(\lambda,\beta)}$ search. The values for $\tau$ and $w_{i,(\lambda,\beta)}$ are indicated. Statistical significance is indicated by (*) for a p-value $<0.05$, and (**) for a p-value $< 0.01$ with respect to the CNN prediction.} 
			\label{tab:update_dsc}
			\centering
			\resizebox{\textwidth}{!}{%
			\begin{tabular}{l | c | c | c }
			\hline 
			Task  &     CNN      & GCN \citep{bib:soberanis20}             &  GCN  \\ 
			      &  2D U-Net    & $w_{1,(0.5,1)}$ & updated scores \\
			\hline
			Pancreas & $76.89 \pm 6.6$ & $77.81 \pm 6.3$* ($\tau=0.8$) & $78.20 \pm 6.1$* ($\tau=0.8$, $w_{2,(1,1)}$) \\ 
			\hline
			Pancreas-10 & $52.14 \pm 22.6$ & $54.55 \pm 22.1$* ($\tau=1e-3$)& $55.14 \pm 21.6$*  ($\tau=1e-3$, $w_{1,(0.5,0)}$)\\ 
			\hline
			\hline
			Spleen & $93.17 \pm 2.5$ & $95.07 \pm 1.3$** ($\tau=0.999$) & $95.20 \pm 1.4$** ($\tau=0.5$, $w_{1,(0.5,0)}$) \\ 
			\hline
			Spleen-9 & $78.89 \pm 28.4$ & $81.15 \pm 28.9$** ($\tau=0.999$)& $81.72 \pm 29.0$* ($\tau=1e-3$, $w_{1,(0,1)}$)\\ 
			\hline
			\end{tabular}
			}
\end{table}
\subsection{Deep Insights on Prediction, Expectation, and Entropy}\label{subsec:discussion}

\begin{figure}
\begin{center}
\includegraphics[width=0.92\textwidth]{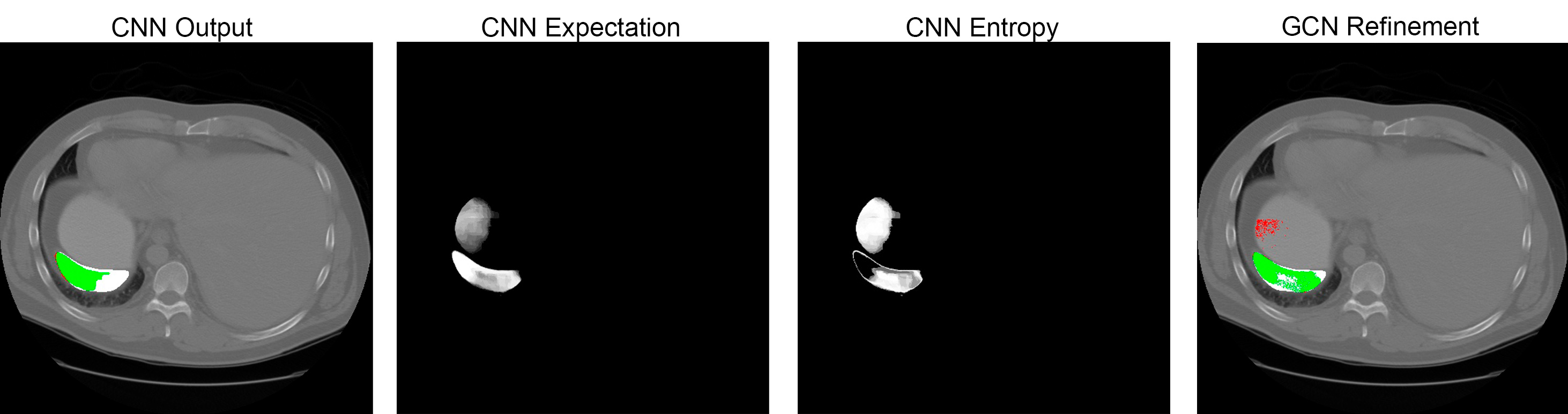}
\end{center}
\caption{Elements used in the graph definition. In the CNN and GCN outputs: Green colors indicate true positives, red false positives, and white false negative regions. For the expectation and entropy: brighter intensities indicate higher values.} \label{fig:graph_components}
\end{figure}

We employed three elements from the uncertainty analysis in the definition of our graph: the CNN's prediction, the CNN's expectation, and the CNN's entropy. Fig. \ref{fig:graph_components} shows an example of these components.

The labels of the graph are given by the CNN's high-confidence prediction. However, from   Fig. \ref{fig:graph_components} we can see that the refinement is similar to the expectation. The expectation is one of the features of the nodes. Also is the main component for the diversity in the edge's weighting function (see section \ref{subsubsec:edges_weighting}). The GCN can learn how to use the CNN's expectation, together with intensity and spatial information,  to reclassify the nodes of the graph. However, it can also generate false positives if the expectation contains artifacts.  Fig. \ref{fig:graph_components} shows an example of this case, where we can see a region in the expectation that does not agree with the ground truth. It can be also noticed that the GCN reduced this region. This can be a result of the random long-range connections included in the graph definition. 

\begin{figure}
\begin{center}
\includegraphics[width=\textwidth]{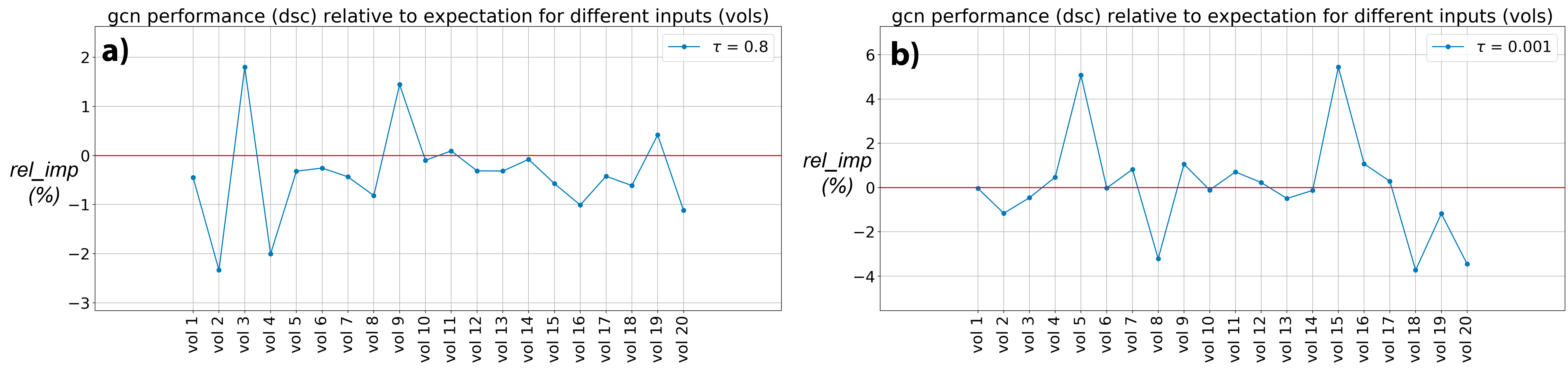}
\end{center}
\caption{Relative improvement (\%) per input volume of different GCN configurations with respect to the expectation of a pancreas segmentation model. The red line indicates the same dsc as the expectation. a) CNN trained with 45 volumes, $\tau=0.8$. b) CNN trained with 10 volumes, $\tau=0.001$.} \label{fig:comparison_expectation}
\end{figure}

In our last experiment, we evaluate the relationship between the expectation and the GCN refinement. For this, we compute the relative improvement between the GCN and the expectation. First, the expectation was thresholded by 0.5. Then we computed its dice score with the ground truth. The relative improvement is computed as:
\begin{equation}
rel\_imp = \frac{gcn_{dsc} - expectation_{dsc}}{expectation_{dsc}} \times 100.
\end{equation}
We compute $rel\_imp$ for every input volume. Fig. \ref{fig:comparison_expectation} shows the results for the pancreas segmentation task, and compares the metric when the expectation was obtained from a model trained with 45 (Fig.\ref{fig:comparison_expectation}a) and 10 samples (Fig.\ref{fig:comparison_expectation}b), respectively, for pancreas segmentation.  

Fig. \ref{fig:comparison_expectation}a shows that most of DICE coefficients (17/20) of the GCN refinement are either below or close to the ones of the expectation. However, three volumes show an improvement in the DICE compared to the expectation. This is different in Fig. \ref{fig:comparison_expectation}b. Here, (13/20) volumes show either better or similar DICE for the GCN compared to the expectation. A possible explanation is that models trained with an adequate number of examples (volumes), their expectation is good enough. In contrast, models trained with a few examples (volumes) have higher uncertainties yielding unreliable expectations. Our results suggest that our GCN refinement strategy is favorable over the expectation or uncertainty analysis in such scenarios.


\subsection{Applicability to other network architectures}\label{sec:other_models}

As a refinement strategy, our proposed method is orthogonal to any segmentation pipeline and can be applied to different CNNs architectures equipped with the MDCO approximation. To verify this, we apply the uncertainty-GCN refinement to the predictions of QuickNat~\citep{bib:roy2019quicknat,bib:roy2019bayesian}, trained on the same two segmentation tasks. The network is trained using the same 45 volumes for the pancreas, and 26 for the spleen, under similar settings as the U-Net. The testing set is the same we used to evaluate the GCN refinement from U-Net. We employed the weighting function $w_{2,(1,1)}$ with $\tau=0.8$ and $\tau=0.5$ for the pancreas and spleen models, respectively.  Table \ref{tab:models_comparison} presents the results for the initial QuickNat prediction and the GCN refinement.  

\begin{table}[h]
			\caption{Average dice score performance (\%) of the GCN refinement compared with the initial CNN predictions. Statistical significance is indicated by (*) for a p-value $<0.05$, and (**) for a p-value $< 0.01$ with respect to the 2D-CNNs predictions.} 
			\label{tab:models_comparison}
			\centering
			\resizebox{\textwidth}{!}{%
			\begin{tabular}{l | c || c | c || c | c   }
			\hline 
			Task  & 3D U-Net & 2D U-Net & Ours & 2D QuickNat & Ours    \\ 
			      & & Initial & GCN-Refinement & Initial & GCN-Refinement  \\ 
			\hline
			Pancreas & $60.14 \pm 10.07 $ & $76.89 \pm 6.6 $ &  $78.20 \pm 6.1 $*  & $61.31 \pm 13.1$ & $61.57 \pm 13.0$  \\ 
			\hline
			\hline
			Spleen & $82.37 \pm 16.8 $ & $93.17 \pm 2.5 $ & $95.20 \pm 1.4$** & $91.83 \pm 6.0$ & $92.97 \pm 2.3$  \\  
			\hline
			\end{tabular}
			}
\end{table}

Reported results show an improvement over the initial CNN model but on a different level compared to the results obtained with the U-Net. While the spleen problem shows an improvement of $1 \%$ over the initial prediction, the pancreas model shows subtle changes. Such differences among different CNNs can be attributed to the epistemic uncertainty inherent to their respective models. In other words, different behaviors of the uncertainty across models can lead to different behaviors in the GCN refinement strategy. However, a deeper analysis of inter-model uncertainty and their relationship with the Graph-based refinement is necessary. 

\subsection{2D vs. 3D Architectures} \label{sec:3d_architectures}

In our experiments, we employed a 2D architecture since it provides us with all the necessary components to test our method. Nevertheless, our refinement strategy is orthogonal to other segmentation approaches and can be applied to any CNN that produces uncertainty measures with MCDO. This also includes 3D models. However, to our best knowledge, MCDO is most commonly employed with 2D models and its translation to 3D might require additional methodological efforts derived from working with 3D architectures together with additional requirements for data-handling due to memory constraints \citep{bib:LaBonte19}. Further, 3D model might not necessarily provide a better initial segmentation compared to a 2D CNN as reported by \cite{bib:yuyin19,bib:Wang19}. In this regard,  we trained a standard 3D U-Net \citep{bib:Oktay18} subdividing the input volume into blocks of $64 \times 64 \times 32$, with a minimum feature size of 16 for the U-Net convolutions. This allows us to use batches of 8 during training on a NVidia Titan Xp of 12 GB. This lead to a dice-score of $60.14 \%$ for the pancreas segmentation problem, and $82.37 \%$ for the spleen segmentation, using the same training, and testing set as the 2D U-Nets (see Table \ref{tab:models_comparison}). As can be seen, under similar conditions, the 3D model presents a lower performance compared with their 2D counterparts. This can be due to the loss of the global context derived from the subdivision of the volume. Fit the entire volume can contribute to the performance. However, this can be limited by the current GPU memory capabilities. Similarly, using a volume-level input will reduce the number of available samples, which can lead to overfitting problems. Even though we are aware of U-Net, V-Net inspired 3D architectures defined for pancreas segmentation that can reach around $80 \%$ of dice score, with the use of data augmentation and auxiliary losses for deep supervision \citep{bib:Zhu17},  or fine-tuning with the addition of attention gates \citep{bib:Oktay18}, we consider that the 2D U-Net gives us good-enough results on both segmentation problems with an appropriate simplicity to evaluate our framework. Nevertheless, investigating our approach using 3D models equipped with MCDO might be an interesting direction.

\section{Discussion and Conclusion}
    In this work, we have presented a method to construct a sparse semi-labeled graph representation of volumetric medical data, based on the output and uncertainty analysis of a CNN model. We have also shown that graph semi-supervised learning can be used to obtain a refined segmentation. We also provided a deep analysis of the weighting function employed to construct the graph. We have shown that diversity is an adequate choice for expectation-based edge weighting. In a similar way, the inverse diversity can also be a good option under certain circumstances. The dependence of the graph to the expectation and the uncertainty analysis method employed could explain the differences in the performance when refining the prediction of two different CNNs. In this regard, alternatives to the uncertainty estimation method, and the use of calibrated uncertainties can be an interesting direction when working with different CNNs models.

\paragraph{Computational Time:} Regarding computational time,  our method requires to define a graph and then train a GCN model. This can require more computational requirements compared with the most efficient versions of CRF. In our experiments, the time required for training and testing the GCN in the constructed graph is around twice the time required for CRF in the fully connected version of our uncertainty graph. This is a time of around 30 sec $\sim$ 1 min for the GCN vs. 13 sec $\sim$ 30 sec for CRF. These numbers do not consider the time for uncertainty analysis and graph construction.

\paragraph{Early Stopping Criteria for the CNN:} Our method is intended for refining a CNN model training with a standard procedure. It is not trained in an end-to-end fashion (jointly with the CNN). Applying an early stopping criterion could have as a consequence an increase in the uncertainty area, generating a bigger ROI, increasing the number of nodes and, hence, increasing the memory requirements for the GCN.

\paragraph{Uncertainty Quantification:} In this work, we have employed MCDO \citep{bib:Kendall17} for the model uncertainty analysis, and found out the expectation could be a good choice for well-trained models, while our GCN refinement shows superior performance, compared to the expectation, in low-data regime. Nonetheless, recently proposed uncertainty measures \citep{tomczack2019learn}, which disentangle the model's uncertainty from the one associated with the inter/intra-observer variability, might be desirable.  

\paragraph{Graph Representation:} We have investigated different connectivity and weighting mechanisms in defining our graph and extracted a couple of features to represent our nodes. However, prior knowledge, e.g. geometry, could be used to constrain the ROI and provide plausible configurations \citep{degel2018domain,oktay2017anatomically}.

\paragraph{Large Organs and Multi-class Segmentation:} We have shown that the model can be applied to different organ segmentation problems and CNN architectures.  Similarly, our results suggest that the performance can depend on the characteristics of the anatomy studied. In this sense, large and stable organs like the liver can derive in performance similar to the spleen. However, further experiments are necessary to verify this. Similarly, large organs can represent a challenge for a graph-based method, since a graph constructed over voxels can lead to high memory requirements. A change on the node representation of the CT data can help in this problem, however, we leave this as future work. In this work, we addressed a binary classification problem. For a multi-class problem, it should be possible to obtain a vectorial expectation representing each class, and the entropy can be computed considering multiple classes. Even though this brings all the elements to formulate the partially labeled graph, given the complexity in the different structures that share intensity similarities between tissues, a different weighting, connectivity, and node features might be necessary to include meaningful information about the anatomies. Similarly, the inclusion of a larger number of structures will lead to a larger number of nodes, making the efficient node representation of multi-class data an interesting future direction. \\


\acks{R. D. S. was supported by Consejo Nacional de Ciencia y Tecnolog\'{i}a (CONACYT), Mexico. 
S.A. was supported by the PRIME programme of the German Academic Exchange Service (DAAD) with funds from the German Federal Ministry of Education and Research (BMBF) when he contributed to this work.}

%
\ethics{The work follows appropriate ethical standards in conducting research and writing the manuscript. This work presents computational models trained with publicly available data, for which no ethical approval was required.}

\coi{The authors declare no conflicts of interest.}





\vskip 0.2in
\bibliography{soberanis20}

\end{document}